# Optimal Climate Strategy with Mitigation, Carbon Removal, and Solar Geoengineering

Mariia Belaia
Harvard John A. Paulson School of Engineering and Applied Sciences
The John F. Kennedy School of Government
Harvard University, Cambridge, MA 02138, USA

**Abstract**

Until recently, analysis of optimal global climate policy has focused on mitigation. Exploration of policies to meet the 1.5°C target have brought carbon dioxide removal (CDR), a second instrument, into the climate policy mainstream. Far less agreement exists regarding the role of solar geoengineering (SG), a third instrument to limit global climate risk. Integrated assessment modelling (IAM) studies offer little guidance on trade-offs between these three instruments because they have dealt with CDR and SG in isolation. Here, I extend the Dynamic Integrated model of Climate and Economy (DICE) to include both CDR and SG to explore the temporal ordering of the three instruments. Contrary to implicit assumptions that SG would be employed only after mitigation and CDR are exhausted, I find that SG is introduced parallel to mitigation temporary reducing climate risks during the era of peak $CO_2$ concentrations. CDR reduces concentrations after mitigation is exhausted, enabling SG phasing out.

**Keywords:** Integrated Assessment Modelling, climate policy, DICE, solar geoengineering, carbon dioxide removal

# 1 Introduction

We need to understand our full potential to limit global climate risk. A wide range of climate policy instruments exists that, combined, equip us with the tools necessary to safeguard the global public good that is a stable climate. These instruments span across different economic sectors and can be market or non-market, private or public, international or regional.

It is useful to view the world as a whole entity, united in the face of changing climate and choosing a climate policy path for the many decades ahead.[1] This viewpoint was conceptualized in the 1990s based on neoclassical economic growth theory in the form of integrated assessment models (IAMs). IAMs aspire to simulate the dynamic interactions between climate and economy on a timescale of a few decades to centuries and to put alternative policy paths into perspective. Traditionally, policy has focused on greenhouse gas (GHG), particularly carbon emissions, as non-pecuniary, negative externalities that economists have used to justify policy intervention. Climate policy thus took the form of emissions abatement paths determined by either cost-effectiveness (CEA) or cost-benefit analysis (CBA).

While emissions control is a necessary instrument of climate policy, it does not fully capture the entire universe of tools for managing climate risk at the global scale. First, it is possible to remove carbon dioxide on a large-scale. While the costs and environmental side effects of Carbon Dioxide Removal (CDR) are uncertain, including CDR as a policy instrument alters climate policy in fundamental ways by breaking the otherwise monotonic relationship between emissions and climate risk.

---

[1] Although a single decision-maker framework diverges from the reality with many heterogeneous actors of different interests, it allows one to articulate the vision and state the aim for the international climate policy negotiations.

Likewise, it is possible to directly alter the radiative forcing (RF) of climate through technical means - a process often called solar geoengineering (SG). While the risks and effectiveness of SG are uncertain, it is likely that global RF can be altered at a relatively small direct cost, weakening the direct link between atmospheric GHG concentrations and climate change.

The bottom line is that both CDR and SG can reduce the (economic) burden that climate change poses beyond traditional emissions control efforts. A key question therefore relates to the timing of these policy instruments. When ordered in terms of acceptability or desirability, I suspect that most participants in climate policy debates put mitigation first, CDR second, and SG third (if including at all). Assuming that they also view this as an appropriate time sequence is natural. In this paper, I challenge this assumed ordering. For this, I conduct an integrated assessment of mitigation, CDR, and SG that form a climate policy portfolio.

I introduce CDR and SG into IAM following the Occam's razor principle of modelling: parsimonious yet congruent with reality, insightful yet manageable. First, I account for strong ties between mitigation and CDR, whereby CDR simply prolongs the mitigation curve beyond zero emissions. Second, SG directly alters RF at zero direct cost (on a scale of GWP). Still, SG indirect costs are deeply uncertain. Following Moreno-Cruz and Keith (2013), I set these damage costs to be as large as the damages from doubling atmospheric $CO_2$ concentration when it is used at a level large enough to counter doubled atmospheric $CO_2$ concentration.

Existing IAM studies generally treat CDR and SG in isolation (Bahn et al., 2015; Bickel, 2013; Bickel and Agrawal, 2013; Chen and Tavoni, 2013; Emmerling and Tavoni, 2018; Goes et al., 2011; Heutel et al., 2016b, 2018; Keith et al., 2006; Marcucci et al., 2017; Rickels et al., 2018). In contrast, I argue that considering mitigation, CDR and SG simultaneously reveals important synergies. As such, CDR allows for SG phasing-out, while SG reduces mitigation and CDR costs. Beyond IAMs, simultaneous treatment of CDR and SG is featured in two simple analytical frameworks (Heutel et al., 2016a; Moreno-Cruz et al., 2017). However, these are silent on the time evolution of a policy mix.

Overall, my modelling results suggest that in the optimal climate policy SG is introduced in parallel with mitigation and prior to CDR. While mitigation is essential for the reduction of global climate risk, CDR and SG separately can increase global welfare. Adding both instruments simultaneously raises the welfare even further by reducing climate policy costs and climate damages from the beginning.

In the following section, I describe CDR and SG and discuss their parsimonious modelling as part of integrated assessment in greater detail. Next, I present the model design and calibration. Section 4 is dedicated to model results complemented by sensitivity analyzes. In Section 5, I draw conclusions.

# 2 CDR, SG, and their treatment in integrated assessments

2.1 Carbon dioxide removal (CDR)

CDR comprises a wide range of technologies that could allow net removal of carbon dioxide from the atmosphere. Several recent publications, using the term negative emissions technologies (NETs), have provided detailed reviews of a range of approaches to CDR (Fuss et al., 2018; Minx et al., 2018; National Academies of Sciences, 2018; Nemet et al., 2018, p. 3; Smith et al., 2016). These include afforestation and reforestation (AR), soil carbon sequestration (SCS), biochar (BC), bioenergy with carbon capture and storage (BECCS), direct air capture with carbon capture and storage (DAC-CCS), enhanced weathering and ocean alkalinization (EW), and ocean fertilization (OF).

Despite CDR's recent salience in climate policy debates, remarkably little research on CDR technologies has been conducted. As a result, current knowledge on CDR is rather fragmented. Biological CDR methods involving forestry and land-use management have the deepest research base and the longest history in climate-motivated research and policy-analysis. Much of the available literature on non-biological methods consists of overviews or meta-analyses that rely on a remarkably small set of primary literature addressing the costs, scalability, and risks of these technologies. Ignorance about abiotic CDR is high, in part, because government support for research on the technologies has been minimal.

Minx et al. (2018, p.) point to the "blurry boundaries" between mitigation and CDR, particularly given that the traditional definition of mitigation includes human intervention to enhance the sinks of greenhouse gas emissions. As such, with many of the land-use based technologies, the distinction between reducing emissions and enhancing sinks can be fuzzy. Some CDR technologies, such as ocean iron fertilization, would indeed be substantially decoupled from mitigation, but many have strong technical and/or policy linkages with mitigation. For example, bioenergy and DAC can be used in the absence of CCS to produce electricity or liquid fuels with very low lifecycle $CO_2$ emissions (Blanco et al., 2018; Keith et al., 2018; Zeman and Keith, 2008). It is through the use of geological storage that bioenergy and DAC truly become negative emissions technologies. BECCS is most often considered as an integral part of electric sector CCS, particularly because the co-firing of coal and biofuels can be a cost-effective way to manage the variability of biomass supply. Indeed, commercial-scale BECCS, in practice, has operated for decades, where $CO_2$ from ethanol plants is sold into the $CO_2$ pipeline system for Enhanced Oil Recovery in the US Permian basin. Additionally, the combination of BECCS with fossil or renewable power is a potentially important pathway for making hydrocarbon fuels with very low life-cycle $CO_2$ emissions. Similarly, DAC is being developed, in part, to provide industrial $CO_2$ to synthesize carbon-neutral hydrocarbon fuels that would provide energy-dense energy-carriers for transportation using primary energy from a carbon-free source, such as wind, solar, or nuclear power.

With regard to modelling CDR as part of global climate policy, it is important to distinguish between detailed-process (DP) and benefit-cost (BC) IAMs (Weyant, 2017). Although both types are used to analyze climate policy, DP IAMs operate at disaggregated sectoral and regional levels, whereas BC IAMs represent both climate and economy in a highly stylized fashion. The latter, however, more explicitly include climate-change feedbacks on the economy.

To date, DP IAMs have focused on one or two specific technologies, but never a full CDR technologies portfolio due, in part, to fragmented knowledge. The majority of DP IAMs consider BECCS to be the only negative-emissions technology (Azar et al., 2010; Kriegler et al., 2013; Luderer et al., 2013; van Vuuren et al., 2013), although some also include reforestation/afforestation (Humpenöder et al., 2014; van Vuuren et al., 2018). Integrated assessments including more than one CDR include Chen and Tavoni (2013) and Marcucci et al. (2017) (DAC and BECCS) and Strefler et al. (2018b) (DAC-CCS, BECCS, and re/afforestation). Meanwhile, expanding the portfolio would allow hedging against potential risks of large-scale deployment of a single technology, while increasing overall negative emissions potential (Fuss et al., 2018).

Among the most well-known BC IAMs, PAGE and FUND do not allow for negative emissions. The more recent versions of DICE allow for net negative emissions (after the year 2150), thus implying some form of CDR. However, this is not made explicit, with single cost curve representing both traditional mitigation and CDR. Thereby, it neither specifies the CDR technology nor mitigation-CDR portfolio composition over time.

Only a few examples exist in which these types of models have explicitly considered CDR. Keith et al. (2006) use a version of the DIAM model that has been extended to include DAC, as well as a zero-carbon backstop technology, as policy instruments quite distinct from mitigation and with separate cost functions. Similarly, Rickels et al. (2018) extend DICE to include a generic version of CDR as a separate policy option with a distinct cost function, which does not alter over time.

Meanwhile, the strong linkages with mitigation discussed above should be reflected in the way CDR is incorporated in BC IAMs. For this, let us consider two extreme approaches to modeling CDR in relation to mitigation in BC IAMs: *independent* and *integrated*. In the former, CDR is a separate technology with its own supply curve that is unrelated to mitigation. In the latter, CDR simply extends the mitigation curve beyond zero emissions. Under this approach, no distinction between CDR and mitigation exists when net emissions are positive. Of course, the full approach would disaggregate both mitigation and CDR into a complex network of technologies linked by energy demands, supply chains, materials requirements, and technological learning. My argument is that the *integrated* assumption is a closer approximation to the complexities found in a fully disaggregated model than is the *independent* assumption.

## 2.2 Solar geoengineering (SG)

SG counteracts anthropogenic GHG radiative forcing (RF) either by reducing the incoming solar radiation, or by increasing planetary albedo, thereby cooling the Earth. Natural parallels, albeit inexact, to SG include the cooling caused by the emissions of aerosols and particulates from volcanic eruptions and from human activities such as fossil fuel burning. As such, since the Industrial Revolution, aerosols are estimated to have created a total effective RF of -0.9 W/m$^2$ with 5% to 95% uncertainty between -1.9 W/m$^2$ and -0.1W/m$^2$ (van Vuuren et al., 2018).

SG does not alter GHG concentrations, but it weakens the link between atmospheric concentrations of GHGs and negative impacts of climate change. In this sense, it resembles climate change adaptation. Yet, unlike adaptation, SG is non-excludable and non-rivalrous, and it is not naturally limited to a specific place or region. That is, SG is a global public good.

Furthermore, while global warming occurs slowly in response to increased atmospheric GHG concentration, climate response to a decrease in incoming solar radiation is rapid (Myhre et al., 2013). Thus, SG allows for more flexibility, be it deferring the deployment or adjusting policy in the process.

Modelling SG in global IAMs boils down to two fundamental components: the amount of SG and its cost. A number of IAMs model the amount of SG in terms of the aerosol injection rate, i.e. TgS per year (Bahn et al., 2015a; Emmerling and Tavoni, 2017; Gramstad and Tjøtta, 2018; Heutel et al., 2018), whereas others directly introduce SG as changes to RF (Goes et al., 2011; Moreno-Cruz and Keith, 2013). I argue that SG is best articulated in IAMs in terms of RF because modeling deployment strategies goes beyond simple climate models used in IAMs, whereas the forcing efficiency varies across deployment strategies (Dai et al., 2018).

SG costs comprise both direct (deployment) and indirect ("side effects") costs. The estimated engineering costs for annual delivery of albedo modification material to the altitude of 20-30 km range between 1-3 billion 2010 USD for 1 Mt/year of material delivered and between 2-8 billion 2010 USD for 5 Mt/year (McClellan et al., 2012). These are not the only costs to be considered in thinking about SG, however. SG does not "magically" reverse climate change induced by increased atmospheric GHG concentrations. While it lowers temperature, it does not address $CO_2$-specific impacts such as ocean acidification, and it potentially introduces new risks (McClellan et al., 2012). These new risks, while highly uncertain, are dependent upon the SG strategy (MacMartin et al., 2013).

Since the direct costs of SG are negligible on a scale of gross world product (GWP), without loss of generality, it is natural to set it to zero. Although some studies chose to explicitly introduce deployment costs specific to stratospheric aerosol injection, they recognize the insignificance of the role it plays (Eric Bickel, 2013). SG indirect costs are deeply uncertain and need to be accounted for. Current IAMs that include SG are well aware of this, which is reflected in either sensitivity analyses (Gramstad and Tjøtta, 2018) or full uncertainty treatment (Emmerling and Tavoni, 2018).

A few studies have attempted to prescribe certain SG time-paths and to analyze their implications (Bickel and Lane, 2009.; Goes et al., 2011). These time-paths are generally too long for meaningful forecasts and decades-long fixed-SG scenarios go against the responsive nature of SG. Here, I call for more realistic time-paths that recognize SG as a policy choice with costs and benefits that are dependent upon its implementation (Keith and MacMartin, 2015). That said, none of the existing IAM studies depict SG as a multi-dimensional control tool that allows for the tailoring of its implementation to balance the side effects with the benefits.

## 3  The Model

I build on the 2016-R2 version of the Dynamic Integrated model of Climate and Economy (DICE) (Nordhaus, 2018). DICE emerged in 1989 as the first attempt to compute GHG emissions trajectories for optimal global climate policy. A summary of the DICE theoretical framework is given in Appendix A. Briefly, it employs a Pigouvian tax framework to set a price on emissions. Model simplicity facilitates interdisiplinary communication, while transparency and the fact that it is well-documented warrants DICE's extensive use. Its applications have gone beyond academic literature: the model is used to inform climate policy by providing the marginal monetized carbon externality value - the social cost of carbon (Interagency Working Group on Social Cost of Carbon, 2015).

This paper's extensions to DICE consist of introducing CDR and SG, and recalibrating the model from a 5- to a 1-year timestep in order to facilitate the representation of the fast-acting nature of SG. The latter is described in the Appendix B.

*Carbon dioxide removal (CDR)*

Mitigation in DICE is represented by μ, the fraction of baseline "industrial"[2] emissions that are "under control"; realized emissions are the baseline emissions mulitpled 1- μ. Total mitigation costs per unit of gross output, Λ, are a non-linear function of the mitigation level,

$$\Lambda_t = \theta_{t,1}\mu_t^{\theta_2},$$

where $\theta_2$ is set to 2.8, indicating a convex marginal cost function. $\theta_{t,1}$ is tied to the cost of a zero-carbon technology; note $\Lambda_t=\theta_{t,1}$ for $\mu_t=1$. This cost is assumed to decline over time. Initially the upper bound on $\mu_t$ is set to 1, but it is allowed to rise to 1.2 after 2150, which Nordhaus (2018) explain as follows: "it is assumed that there are no 'negative emissions' technologies initially, but that negative emissions are available after 2150." Thereby, it is implicitly assumed that the cost of first unit of carbon removed is equal to the cost of last unit mitigated.

In scenarios not including CDR, I keep the upper bounds on μ at 1 for all periods; in scenarios including CDR, I allow for μ>1 at any time (and without an upper limit) because some kinds of CDR exist today. My treatment is consistent with the integrated approach described in Section 2. Figure 1 illustrates the joint marginal cost curves for mitigation and CDR, showing the decline over time.

---

[2] Industrial emissions represent all carbon emissions from energy use and are related to gross output.

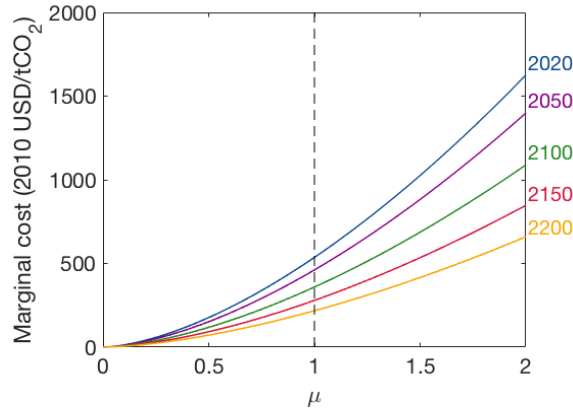

Figure 1 Marginal cost of abatement and CDR, 2010$/tCO2

While in reality CDR begins earlier than full mitigation takes place, traditional mitigation would remain the largest share in a cost-effective mitigation-CDR portfolio as long as its marginal costs are below marginal costs of CDR, or until full mitigation is reached. Furthermore, CDR per se constitutes a portfolio of heterogeneous technologies that enter policy at different times and levels. The integrated modelling approach allows me to abstract from these complexities and focus on the big picture instead. Since net negative emissions clearly indicate the use of CDR, I reserve the term CDR for net negative emissions in later discussions.

*Solar geoengineering (SG)*

I incorporate SG in DICE by allowing the decision maker to choose a level of SG in terms of negative radiative forcing (RF), given by the term $F_t^{SRM}$. Accordingly, I modify the RF equation by substracting the term $F_t^{SRM}$:

$$F_t = F_{2 \times CO2}(\log_2(\frac{M_t^{at}}{M_{1750}^{at}})) + F_t^{ex} - F_t^{SRM}.$$

In line with Moreno-Cruz and Keith (2013), I assume that the side effects increase quadratically with SG, and following Goes et al. (2011), I represent them proportional to the ratio of SG forcing to the forcing from the doubling of atmospheric CO₂:

$$D_t^{SRM} = d^{SRM} \cdot (\frac{F^{SRM}}{F^{2xCO2}})^2.$$

The side effects of SG then add to the climate change damage costs. No real empirical/theoretical evidence exists to calibrate $d^{SRM}$, with current studies suggesting that SG side effects are rather small (Kravitz et al., 2014; Irvine et al., 2019). Given the current perception of, and uncertainty around SG, however, I make the default asumption that damages are large. Specifically, following Moreno-Cruz and Keith (2013), I assume that economic costs induced by the amount of SG required to offset warming from 2×CO₂ are the same as the associated warming in the absence of SG. Thereby, I set $d^{SRM}$, to 2.2% of GWP.

## 4 Results

### 4.1 Default Calibration

To begin, I evaluate two climate policy portfolios: mitigation+CDR and mitigation+SG. I compare and contrast these with baseline and mitigation-only policies. Following Nordhaus (2018), the baseline policy assumes that the 2015 level of mitigation is sustained over the next decades (3% of business-as-usual emissions under control). Under mitigation-only policy, neither negative emissions nor SG are considered.

The time evolution of the portfolios and their implications are illustrated in Figure 2. When climate policy is limited to mitigation, zero emissions are reached by the year 2120, thereby stabilizing atmospheric carbon concentration (as reflected in the RF). These concentrations decline over time due to natural sinks, albeit very slowly. As a result, the global mean temperature is stabilized at 4°C for the full length of the simulation.

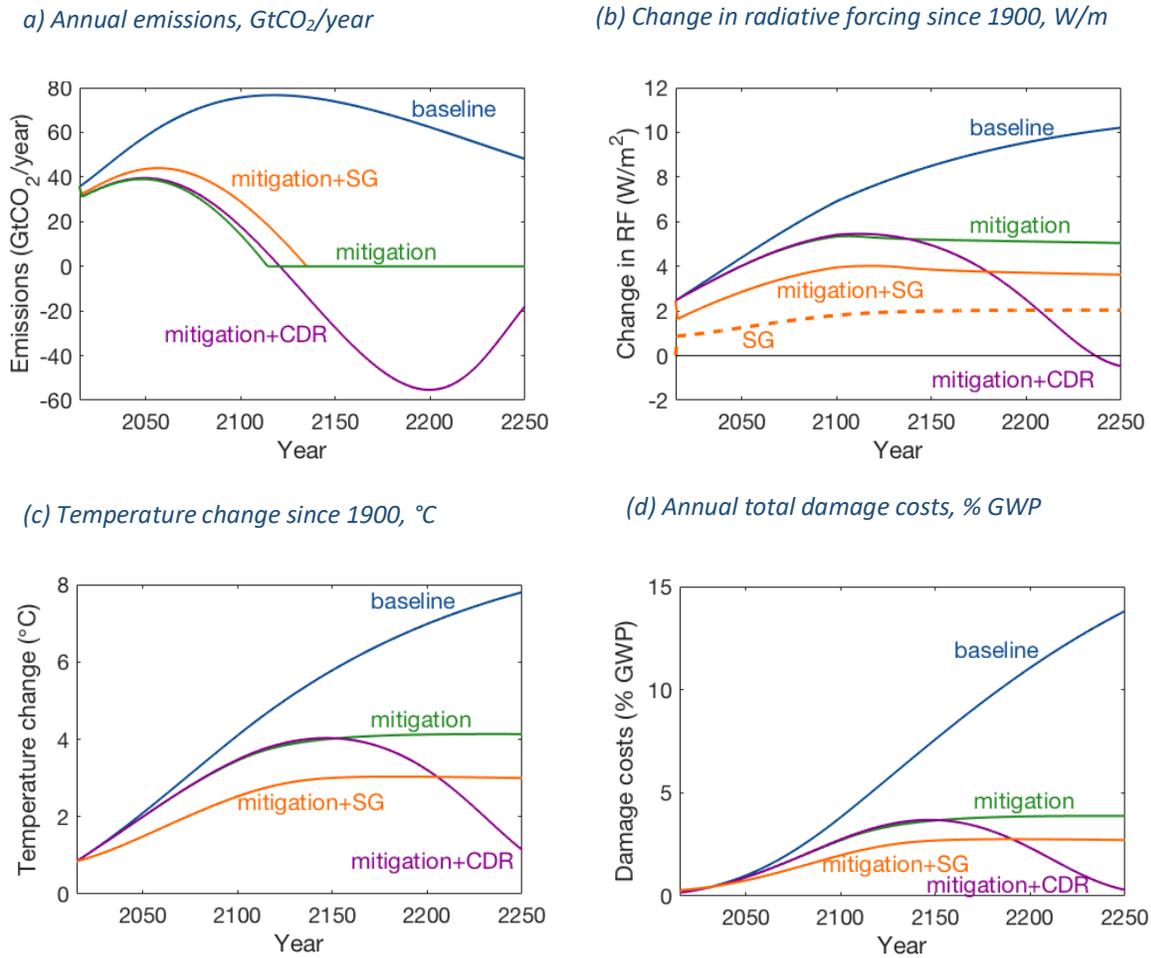

Figure 2 Model results. Selected variables for four alternative climate policy portfolios: baseline (blue), mitigation-only (green), mitigation & CDR (purple) and mitigation & SG (orange). Dashed line depicts the absolute value of SG RF.

With SG added to the portfolio, and in view of perfect foresight, emissions follow a path similar to the mitigation-only case but with a slightly higher and later peak, and a later date of reaching zero emissions. SG begins immediately and follows the dynamics of atmosperic carbon concentration. Hence, its persistence for the whole simulation period at a level that is approximately half of GHGs RF. Given the negative RF from SG, the temperature stabilization level is decreased to 3°C, with lower damage costs (even including those from SG itself). Overall, SG acts to reduce both mitigation costs and climate change damages with many of the benefits realized immediately. As such, it reduces the optimal carbon tax (and the social cost of carbon (SCC)) by 30% already in 2030. Although SG reduces the level of SCC, the latter keeps rising with time (Figure C1, Appendix).

(a) Percentage change in BGE relative to baseline      (b) Side effects and benefits of SG in 2050, % GWP

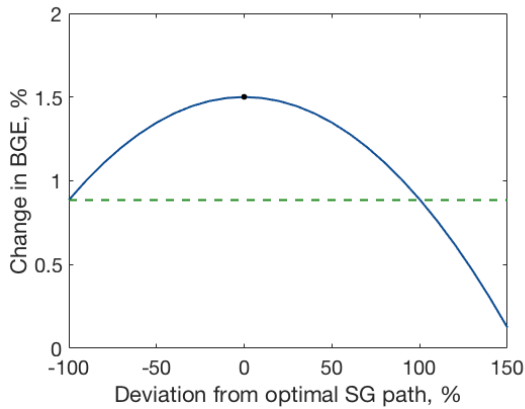 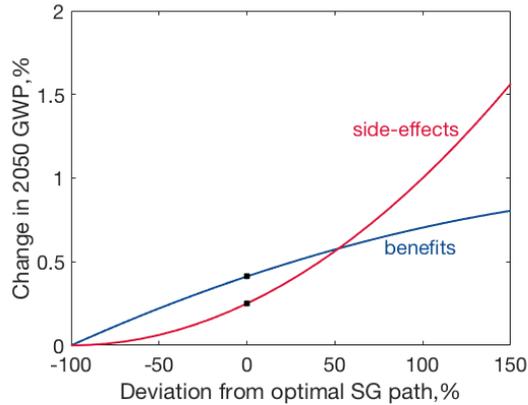

Figure 3 Change in BGE relative to the baseline policy (a) and costs and benefits of SG realized in 2050  (b) following deviation from the optimal level of SG. Black dotes identify the location of the optimal solution. Green dashed line in (a) depicts the level that corresponds to the mitigation-only policy.

Yet, optimal SG is moderate. Figure 3 demonstrates the reason SG is not used to further reduce temperature changes. For this, from the optimal mitigation+SG policy, I fix the mitigation path and perturb the SG path. Tracking the associated changes to the balanced growth equivalent

(BGE)[3], it is evident that an overuse of SG will not only decrease its benefits, but also counteract some of the benefits from mitigation (Figure 3a). The underlying rationale is revealed in Figure 3b, which shows the SG benefits and side effects realized in the year 2050. The SG benefits constitute avoided climate damages, calculated as the difference in climate damages between scenarios with and without SG. Since climate damages grow superlinearly with respect to temperature (and RF), the marginal SG benefits are diminishing. SG side effects, on the other hand, grow superlinearly. Hence, increasing amounts of SG are associated with diminishing net benefits, which eventually become negative.

Let us now return to Figure 2 and direct our attention to the mitigation+CDR portfolio. In view of perfect foresight, the inclusion of CDR slightly delays the mitigation efforts relative to mitigation-only policy. As such, the optimal carbon tax in 2030 is lowered by 3%. Unlike SG, however, over the long term, CDR brings down atmospheric $CO_2$ concentrations and returns global temperature to its pre-industrial level. Therefore, CDR decreases SCC over time (Figure C1b).

---

[3] Balanced growth equivalent (BGE) is a commodity measure of welfare first introduced by Mirrlees and Stern (1972).

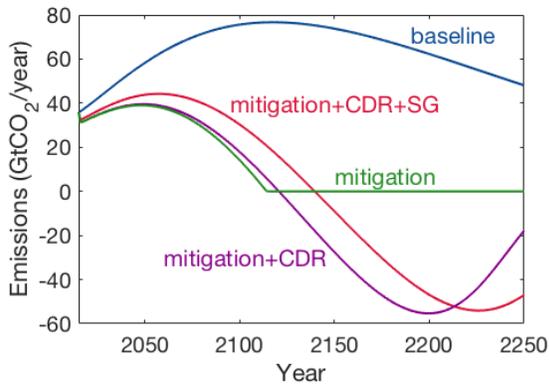
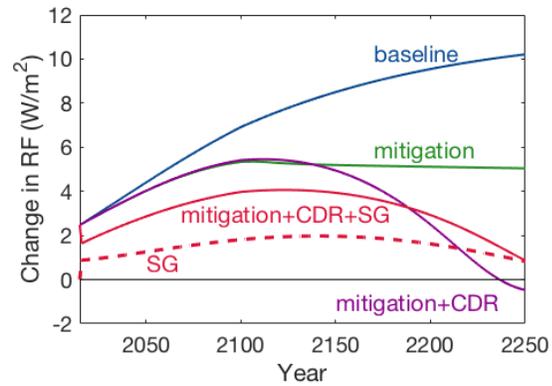
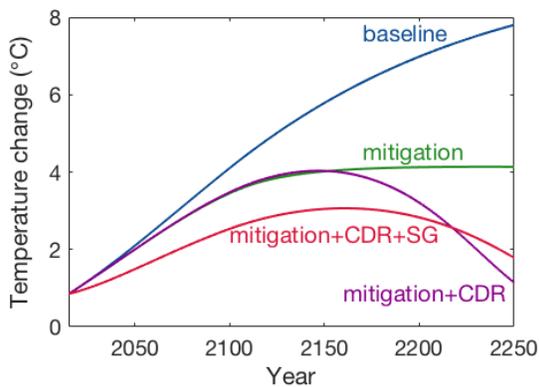
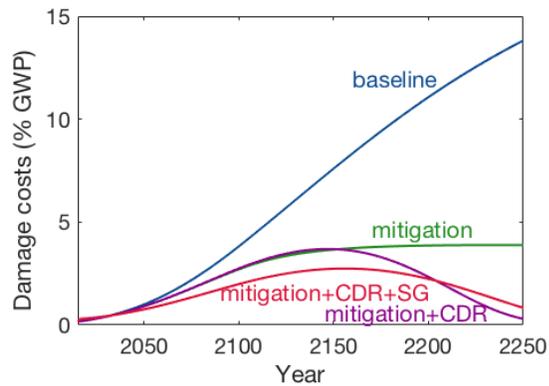

*Figure 4 Results for the default calibration. Selected variables in three alternative climate policy portfolio compositions against the baseline (blue): mitigation-only (green), mitigation+CDR (purple) and mitigation+ CDR+SG (red). Red dashed line identifies the absolute value of SG RF.*

Next, I combine all three instruments in one climate policy portfolio (Figure 4). SG enters policy parallel with mitigation and before CDR is used to reach net-zero emissions. The inclusion of CDR allows for the gradual phasing-out of SG over time. Hence, SG peaks around the time CDR is used to reach net-zero emissions. Since comparative advantages differ across these instruments, the largest BGE value is reached when all three are considered (Figure D7). The "full" portfolio reduces global climate risk at lower near- to medium-term carbon tax (Figure C1a). It ensures long-term decline in damages and SCC (Figure C1b) via CDR, where their peak values are reduced via SG.

## 4.2 Sensitivity analysis

Here, I test the robustness of the findings presented in the previous section to model assumptions, both inherited from DICE and those I make. Table I summarizes the results of sensitivity runs with associated figures and scenario descriptions available in the Appendix D. First, I confirm the robustness of the policy instruments' time sequence: SG enters before and peaks around the time when CDR is used to reach net negative emissions. Second, I confirm the role of CDR as a guarantee of SG phasing-out and the role of SG in reducing near-term climate damages and optimal carbon tax. Third, improvements in the balanced growth equivalent from the baseline scenario are largest when all three instruments are considered (Figure D7).

*Table I The mitigation+CDR+SG portfolio: sensitivity results*

| Scenario | Peak annual emissions | | Reaching net-zero emissions, year | Peak annual CDR | | Cumulative industrial emissions TtCO$_2$ | Peak SG | | Peak temperature change | |
|---|---|---|---|---|---|---|---|---|---|---|
| | year | % base | | year | GtCO/year | | year | W/m$^2$ | year | °C |
| Default | 2057 | 72 | 2142 | 2228 | 55 | 4.0 | 2143 | 1.98 | 2163 | 3.1 |
| S1. Larger discount rate | 2071 | 78 | 2183 | 2294 | 59 | 6.0 | 2188 | 2 | 2201 | 3.9 |
| S2. Zero utility discount rate | 2037 | 64 | 2102 | 2174 | 41 | 2.0 | 2100 | 1.8 | 2128 | 2.1 |
| S3. 2 × climate damages | 2049 | 69 | 2124 | 2200 | 50 | 3.1 | 2122 | 2.7 | 2143 | 2.1 |
| S4. Linear SG side effects | 2057 | 69 | 2140 | 2218 | 65 | 4.0 | 2134 | 2.5 | 2104 | 2.8 |
| S5. 2 × SG side effects | 2053 | 71 | 2133 | 2216 | 55 | 3.6 | 2136 | 1.1 | 2156 | 3.5 |
| S6. 0.1 × SG side effects | 2085 | 84 | 2230 | 2369 | 42 | 8.7 | 2218 | 6.8 | 2227 | 1.0 |

In addition, I demonstrate that the need for CDR and SG is independent of the generational discount rate on welfare (Figure D2). Yet, larger discounting implies more emissions, which drives the need for both larger CDR and more SG. Since CDR is a long-term strategy, it is more sensitive to discounting than is SG. As such, by increasing the rate of pure time preference from 0 % to 3% per year, peak CDR grows by 44%, while peak SG increases by 10 % (S1 vs. S2). Further scenarios, S3-S6, demonstrate that the optimal amount of SG and its potential to reduce temperature strongly depend on its benefits and side effects.

### 4.3 The 2°C target framework

Since policy discussions center on reaching a 2°C target, I also present some basic results of attempting to attain this goal using the framework developed here. For this, I remove the damage costs function, substituting it with a temperature constraint i.e. $T^{atm} \leq 2, \forall t$. Note that this moves us from a cost-benefit to a cost-effectiveness approach, which is inconsistent with what is recommended by most economists. Furthermore, the 2°C target lacks a physical basis and is vague regarding the reference point. For these reasons, I emphasize that I do not support the 2°C target framework as a policy guide, even as I present these results.

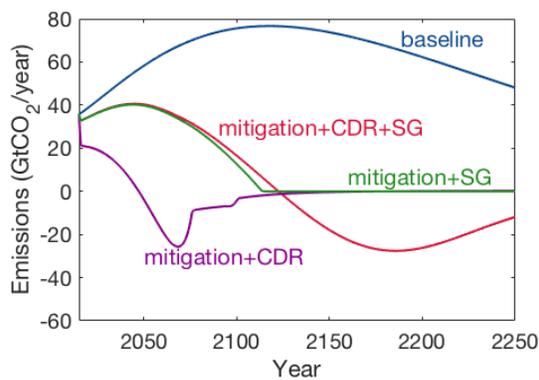
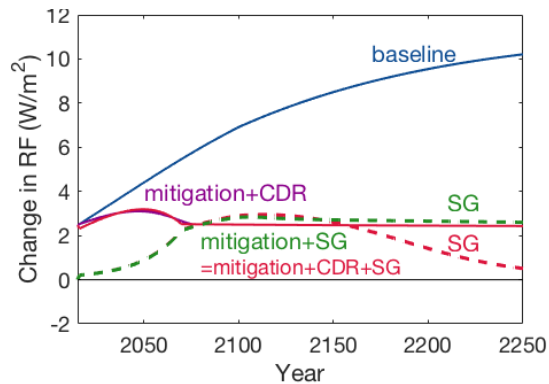
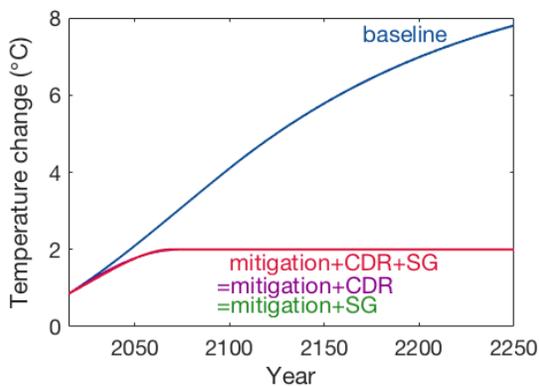
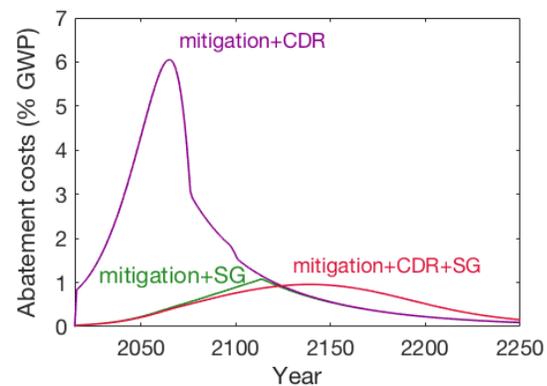

Figure 5 The 2°C target framework results.

In DICE, 2°C is unattainable under the mitigation-only policy. As such, it is not shown in Figure 5. Introducing SG along with mitigation makes the target feasible at a minimum near-term cost, but it requires sustained use. Introducing CDR along with mitigation, on the other hand, already stabilizes atmospheric $CO_2$ concentration (and RF) at a lower level by 2100. This requires reaching net-zero emissions by 2050 and incurring annual costs of up to 6% GWP per year. The combined use of CDR and SG, in addition to mitigation, allows for reaching the target at a lower cost and with only temporary use of SG. Worth noting is that while CDR stretches the carbon budget, SG weakens the link between the carbon budget and the temperature target.

## 5. Conclusions

The present study explores alternative global climate policy portfolio compositions. Not surprisingly, and in line with Moreno-Cruz et al. (2018), I found that the largest gains are associated with a "full" portfolio that includes both CDR and SG, in addition to mitigation. In view of perfect foresight, CDR and SG separately redistribute over time the emissions control that results from a mitigation-only CBA, leading to short-run mitigation efforts that are less stringent. They do so to a degree that allows taking advantage of the decline over time in abatement costs, while keeping the atmospheric carbon stock or climate changes in check. The fast-acting nature of SG, in addition to its low cost, makes it more effective in reducing the near-term optimal carbon tax (and the social cost of carbon). For the default parametrization, SG reduces the optimal carbon tax in 2030 by 30%, whereas for CDR, it is by 3%.

Unlike previous IAMs, I include CDR and SG simultaneously. In contrast to what I expected to be the desired order of policy instruments, I found that the optimal use of SG is before CDR is employed to reach net-zero emissions. Simultaneous treatment also allows for the uncovering of interesting insights about the role of each instrument. First, I found that CDR is best used to reduce long-term climate impacts, as opposed to the currently prevalent framing of CDR as a means to reach the 2°C target by 2100 (Chen and Tavoni, 2013; Kriegler et al., 2013; van Vuuren et al., 2018). The near- to medium-term climate damages and carbon tax are diminished via SG. However, more importantly, I demonstrated that optimal SG is temporary and moderate. Thus, I provide evidence against long-term persistent SG scenarios that appeared in a few IAM studies (Bickel and Lane, 2009; Goes et al., 2011). Finally, I demonstrated the robustness of the instruments' time sequence and their role in climate policy to a range of model assumptions and modeling frameworks. I also provided evidence against the claims that the need for SG and CDR is driven by discounting.

## Acknowledgements

I am extremely grateful to David W. Keith and Gernot Wagner for useful discussions that helped to shape this paper. I thank Michael Funke, Timo Goeschl, Juan Moreno-Cruz, and Dale S. Rothman for insightful comments on an earlier draft. I also acknowledge financial support from the Harvard's Solar Geoengineering Research Program (SGRP).

# Appendices

## Appendix A The DICE model

DICE describes continuous interactions between the economy, welfare, and the climate system in a fully-coupled fashion. It extends the neo-classical Ramsey growth model to include investments in emissions abatement, which compete with consumption and savings for the economic output net of climate change damage costs. While abatement can reduce present day consumption, it allows for greater consumption in later years by reducing future levels of atmospheric carbon, a negative "natural capital stock", and thereby future damage costs. Thus, DICE is useful for comparing alternative climate policies, expressed as emission abatement pathways, and selecting the optimal policy under specific assumptions.

Gross economic output is given by a Cobb-Douglas production function of labor, capital stock, and economy-wide technology. While labour growth and technological change are exogenous, the capital stock is determined by savings, which compete with current consumption and investments in emissions abatement. The influence of economic activity on climate is captured via global emissions, which are proportional to global output and depend upon an exogenous emissions intensity of output and the chosen level of abatement. The former declines over time based on an assumption that the economic system will continue to become more energy efficient and/or that the energy system will naturally decarbonize even in the absence of climate policy. Mitigation costs are given as an exponential function of the level of abatement. These costs also decline over time based on an assumption about the energy-system technological progress. Next, the climate module of DICE describes the logical flows that link emissions to an increase in global mean temperature. It comprises a carbon cycle, radiative forcing and associated changes in climate as represented by global mean temperature, in both atmosphere and oceans. The feedback of climate change on the economy is modelled via a so-called damage function. The basic premise is that damage costs increase quadratically with global mean temperature increase. The function is then calibrated using a two-point fit approach based on the estimates from Nordhaus and Moffat (2017). The damage function includes adaptation costs and residual damages, where adaptation is assumed to be optimal. Although DICE is a highly aggregated global model, its calibration is based upon the country-level data from major countries, considering climate change's differentiated impacts. Aggregation uses exchange rates in PPP terms. The detailed description of the latest version of DICE, DICE2016R2, can be found in Nordhaus (2018)

## Appendix B Finer time resolution

I emphasize that time-varying parameters in DICE2016R2 are calibrated to match the projections. Thereby, I work with the difference equations as oppose to numerical integration of differential equations with finer timestep. The time evolution of exogenous parameters for 5- (*) and 1-year (–) time steps are illustrated in Figure B1. I then recalibrate the carbon cycle model parameters to fit the baseline temperature-change projections. The small difference in the resulting emissions control early-on relates to more responsive decision making (Figure B2).

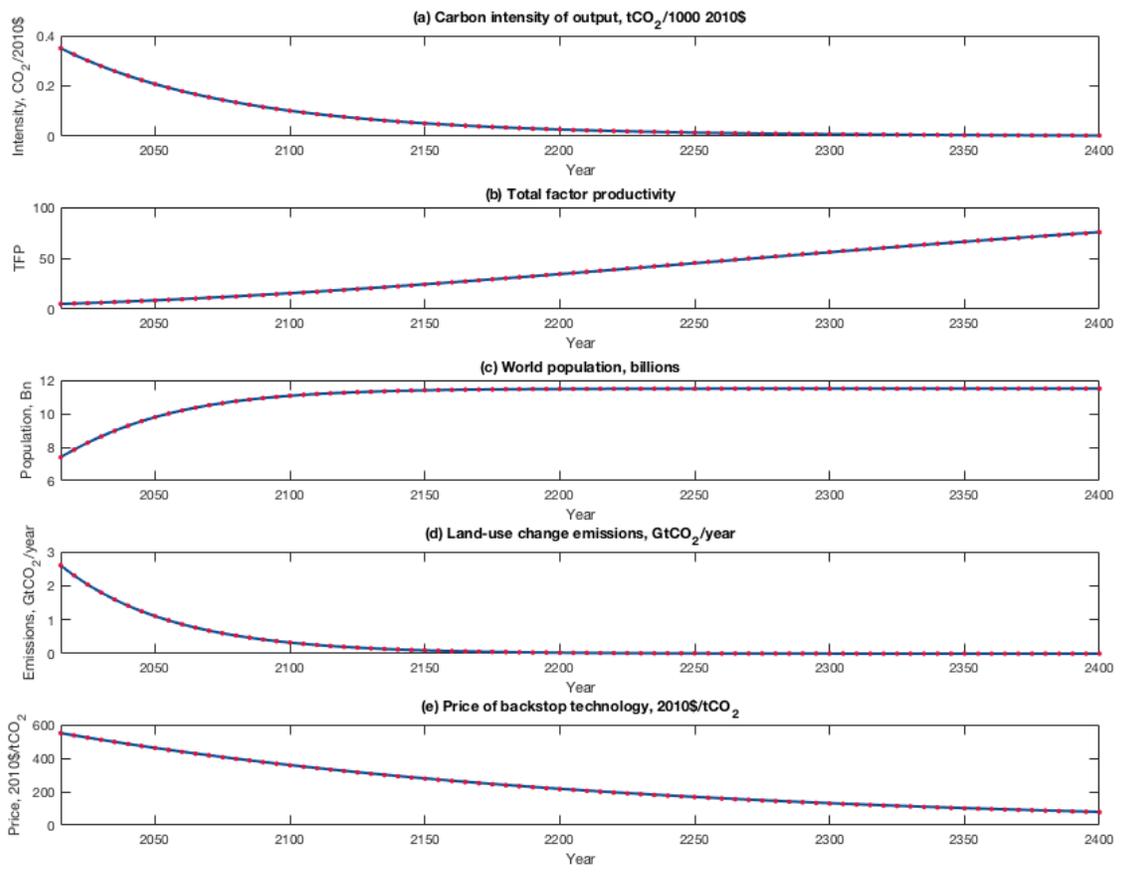

Figure B1 Exogenous time-varying parameters: 1-year time step (blue) and 5-year time step (red).

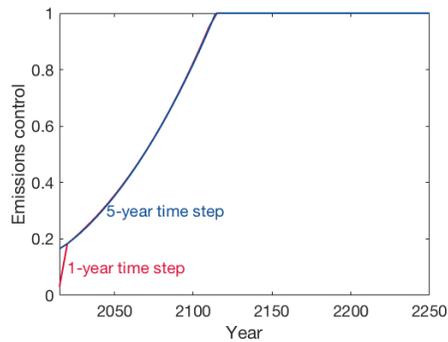

Figure B2 Optimal emissions control in the mitigation-only scenario: DICE2016R2 with 1- and 5-year time steps.

# Appendix C The carbon tax and the social cost of carbon

a) The carbon tax, 2010$/tCO₂    (b) The social cost of carbon, 2010$/tCO₂

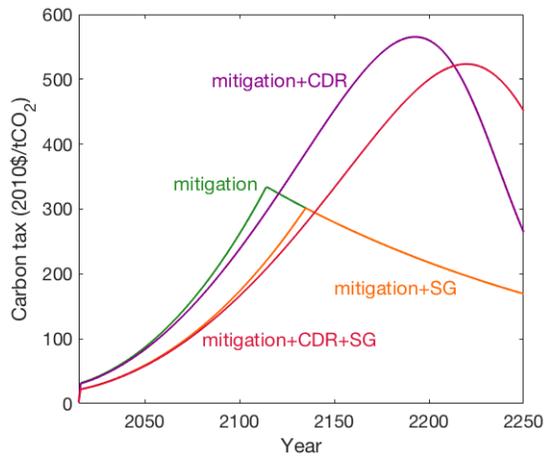 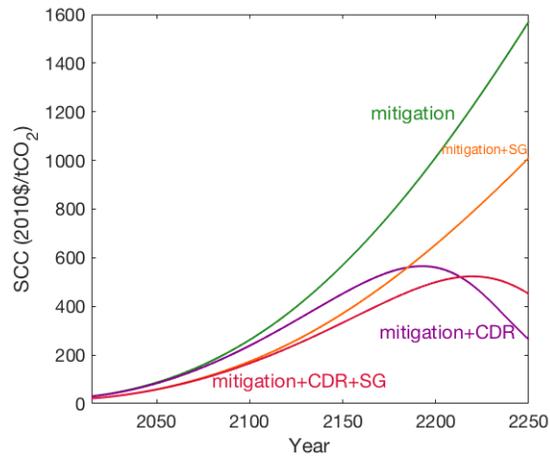

*Figure C1 The optimal carbon tax (a) and the social cost of carbon (b) for alternative climate policy portfolio compositions under default parameters.*

Here, carbon tax is set at the marginal cost of mitigation (or mitigation+CDR, in case CDR is included). The social cost of carbon (SCC) is estimated through the ratio of shadow price of carbon emissions to the shadow price of consumption along the optimized path (Nordhaus, 2017). Thereby, it reflects the welfare impacts of emissions in terms of t-period consumption.

## Appendix D Sensitivity analysis

Table DI lists sensitivity analysis scenarios. These include the social rate of time preference (or utility discount rate) ρ, climate change damage costs $D(T^{atm})$, and SG side effects $D^{SRM}(F^{SRM})$. The results are shown in Figures D1-D6. Figure D7 illustrates the sensitivity of the balanced growth-equivalent (BGE) change to model assumptions, which is presented for alternative

climate policy portfolio compositions.

*Table D-I Sensitivity runs*

| Title | Setting | | Figure |
|---|---|---|---|
| | Default | Scenario | |
| S1. Larger discount rate | $\rho=1.5\%/year$ | $\rho=3\%/year$ | D1 |
| S2. Zero utility discount rate | | $\rho=0\%/year$ | D2 |
| S3. Larger climate change damages | $D(T^{atm}) = a_1(T^{atm})^2$ | $2 \times a_1(T^{atm})^2$ | D3 |
| S4. Linear SG side effects | $D_t^{SRM} = d^{SRM} \cdot (\frac{F^{SRM}}{F^{2xCO2}})^2$ | $d^{SRM}(F^{SRM}/F^{2xco2})$ | D4 |
| S5. Larger SG side effects | | $2 \times d^{SRM}(F^{SRM}/F^{2xco2})^2$ | D5 |
| S6. Smaller SG side effects | | $0.1 \times d^{SRM}(F^{SRM}/F^{2xco2})^2$ | D6 |

*a) Annual emissions, GtCO₂/year*

*(b) Change in radiative forcing since 1900, W/m*

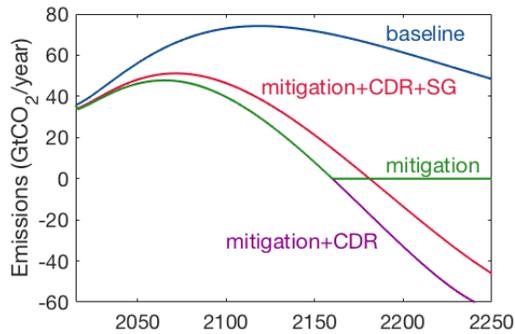
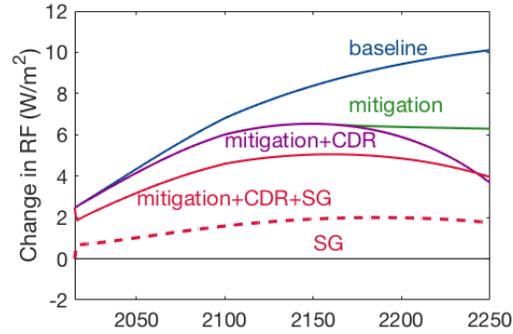

*(c) Temperature change since 1900, °C*

*(d) Annual total damage costs, % GWP*

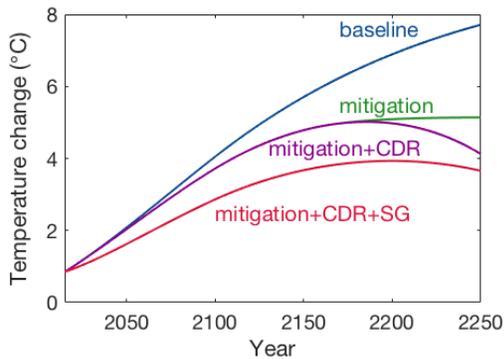
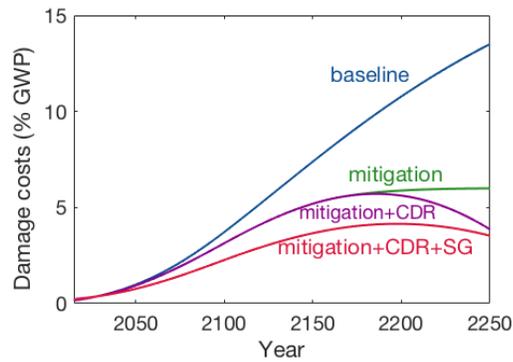

*Figure D1 Larger discount rate.*

*a) Annual emissions, GtCO$_2$/year*  *(b) Change in radiative forcing since 1900, W/m*

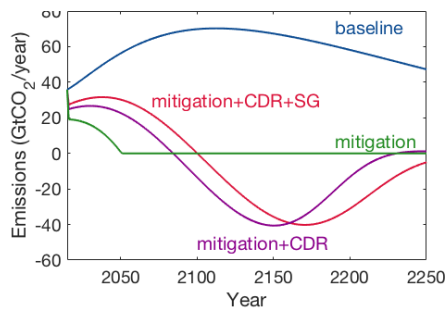 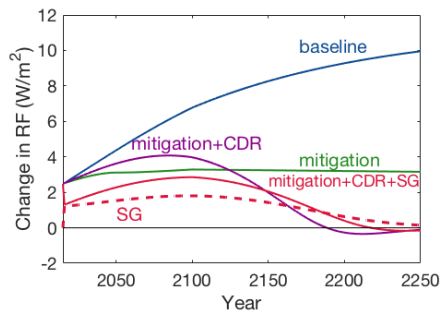

*(c) Temperature change since 1900, °C*  *(d) Annual total damage costs, % GWP*

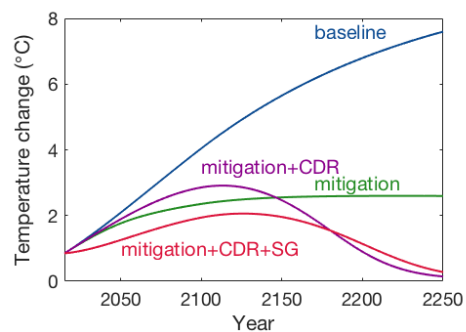 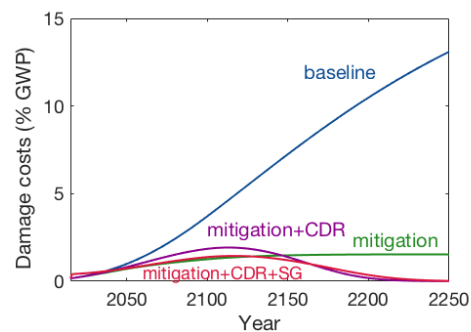

*Figure D2 Zero welfare discount rate.*

*a) Annual emissions, GtCO$_2$/year*  *(b) Change in radiative forcing since 1900, W/m*

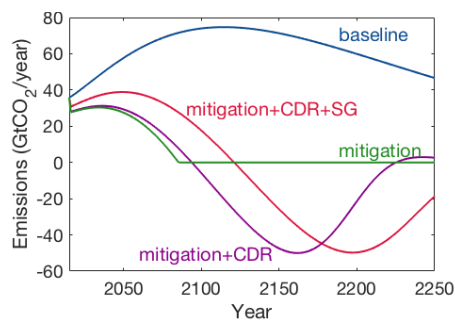 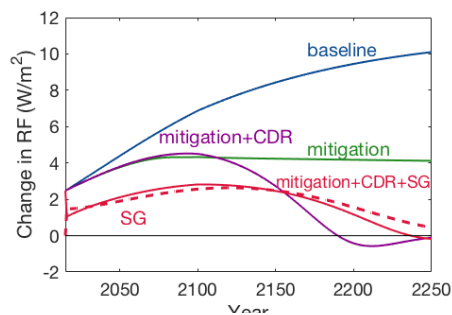

*(c) Temperature change since 1900, °C*  *(d) Annual total damage costs, % GWP*

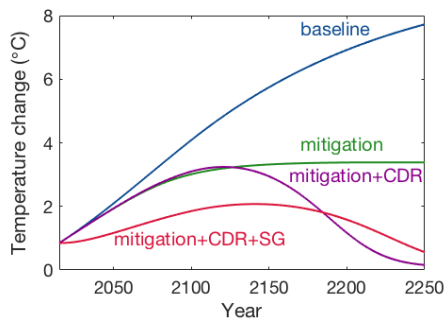 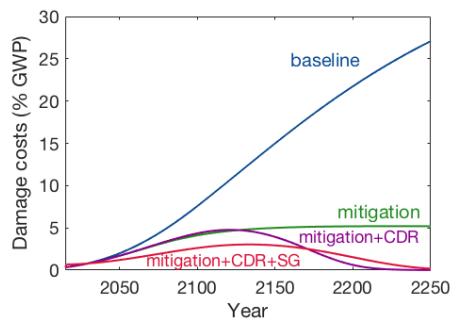

*Figure D3 Larger climate change damages.*

*a) Annual emissions, GtCO₂/year*  *(b) Change in radiative forcing since 1900, W/m*

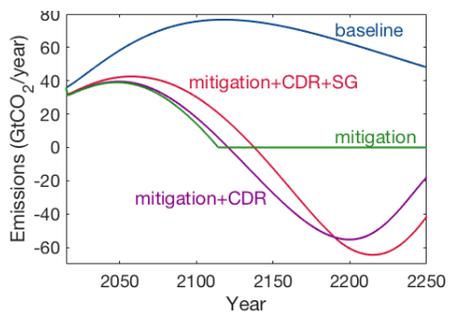 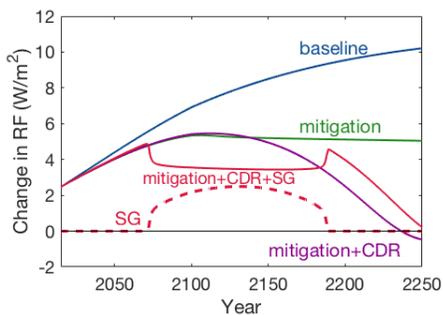

*(c) Temperature change since 1900, °C*  *(d) Annual total damage costs, % GWP*

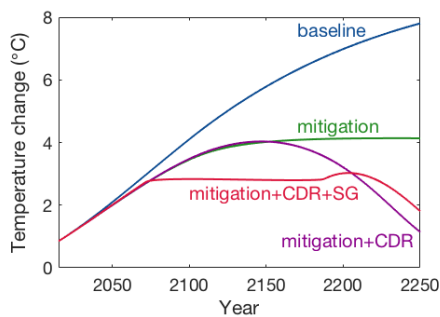 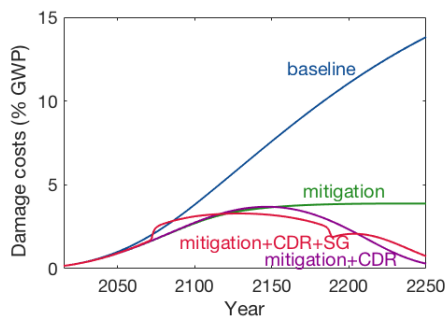

*Figure D4 Linear SG side-effects.*

*a) Annual emissions, GtCO₂/year*  *(b) Change in radiative forcing since 1900, W/m*

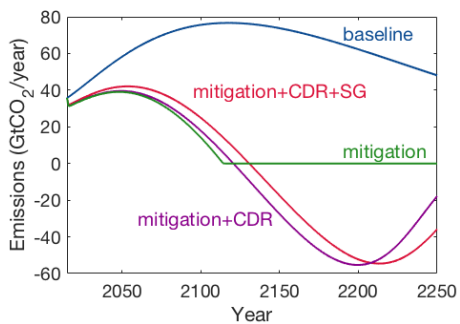 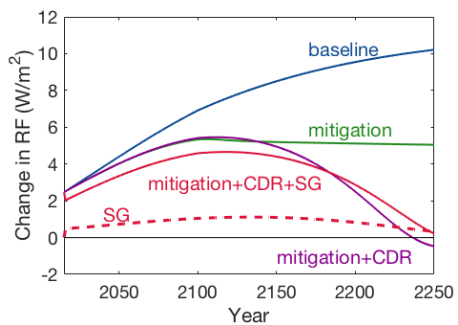

*(c) Temperature change since 1900, °C*  *(d) Annual total damage costs, % GWP*

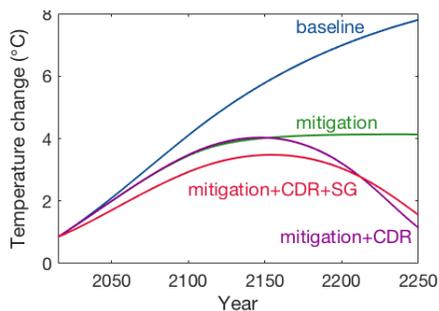 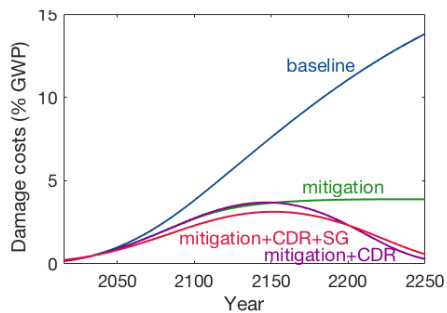

*Figure D5 Larger SG side-effects*

a) Annual emissions, GtCO$_2$/year

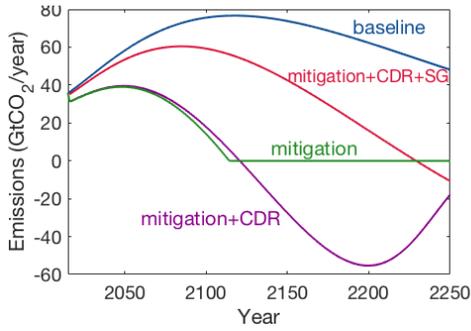

(b) Change in radiative forcing since 1900, W/m

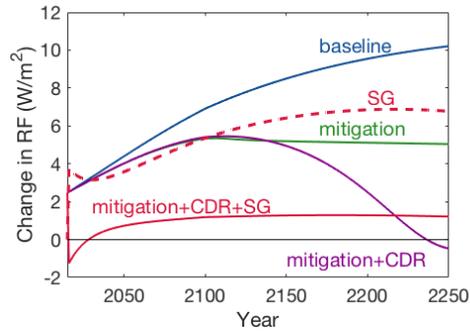

(c) Temperature change since 1900, °C

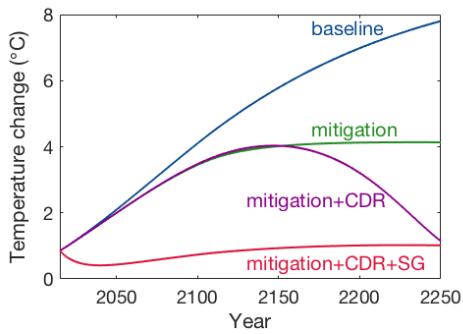

(d) Annual total damage costs, % GWP

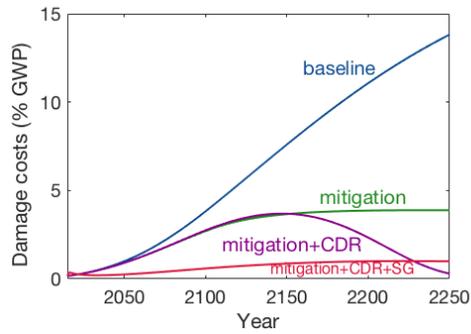

Figure D6 Smaller SG side-effects.

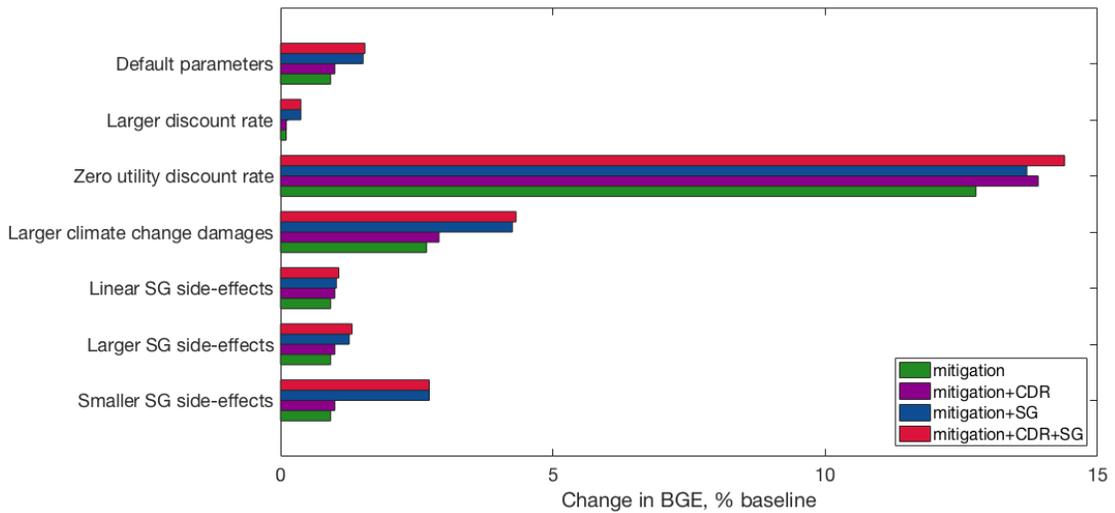

Figure D7 Percentage change in the balanced growth equivalent (BGE) relative to the baseline presented for alternative climate policy portfolio compositions and scenarios.